# Twinning in ferromagnetic Heusler Rh$_2$MnSb epitaxial thin films

Artem Shamardin[1*], Stanislav Cichoň[1], Esther de Prado[1], Jan Duchoň[1], Michal Rameš[1], Ladislav Fekete[1], Jaromír Kopeček[1], Šimon Šťastný[2], Aleš Melzer[2], Michal Hubert[2], Martin Veis[2], Ján Lančok[1], Oleg Heczko[1]

[1]Institute of Physics of the Czech Academy of Sciences, Na Slovance 1999/2, Prague 18200, Czechia

[2]Institute of Physics of Charles University, Ke Karlovu 2026/5, Prague 12116, Czechia

*Corresponding author email: shamardin@fzu.cz

**Keywords:** Heusler alloys; thin film; epitaxy; Rh$_2$MnSb; twinning effect; twin formation; Kerr effect

## Abstract

Epitaxially grown full Heusler alloy of Rh$_2$MnSb thin films were prepared for the first time using DC magnetron sputtering. The films were deposited on MgO [001] substrates with a deposition temperature of 600°C, 700°C, and 800°C. We report the structural, morphological, optical, magneto-optical, and magnetic properties of the films with a 200 nm nominal thickness. The grown-at-600°C film was close to stoichiometric and exhibited $L2_1$ ordering typical for Heusler alloys. The single-phase Rh$_2$MnSb film had a tetragonal structure with lattice parameters close to the bulk material. X-ray photoelectron spectroscopy revealed the metallic character of the film free from contamination. The tetragonal films exhibited discernible regular twinning with the majority of twin domains with the c-axis perpendicular to the surface due to a substrate constraint. The twin formation was studied by atomic force and transmission electron microscopy and by X-ray diffraction. Magnetic measurements showed $T_C$ of about 220-275 $K$ and saturation magnetization of about 55 emu/g, close to the bulk material. Magneto-optical Kerr effect measurements of the film prepared at 600 °C affirmed paramagnetic behavior at room temperature and suggested the half-metallic behavior. The observed properties highlight the potential for further investigations of Rh$_2$MnSb's thin films, focusing on compositional and structural control.

## 1. Introduction

Heusler compounds are series of intermetallic compounds with over 1500 possible combinations. They are classified into two classes based on the stoichiometries of the three constituent elements: 1:1:1 (*XYZ* – half Heusler) and 2:1:1 (*X$_2$YZ* – full Heusler). The prototypes *C1$_b$* and *L2$_1$* with space groups *F-43m* (No. 216) and *Fm-3m* (No. 225) are defined for half and full Heusler, respectively. In a chemical formula, X and Y are transition-metal elements while Z is s-p elements, such as Al, Ga, Sn, Sb, etc. Heusler compounds provide an appropriate research field because of their variety of compositions and flexible characteristics, making them suitable for various applications. Numerous features are present in these alloys, including shape memory effect [1,2], thermodynamic [3,4], half metallicity [5–7], heavy fermion behavior [8–10], magnetic ordering [11,12], and superconductivity [13].

Bulk alloys of the Rh-Mn-Sb system attracted scientific attention half a century ago due to their high ferromagnetic response and Curie temperature [14]. Later, several compounds of this system with specific compositions and crystal structures were identified and characterized [15,16]. In the case of full Heusler Rh$_2$MnSb, the system is reported to exist in a relatively wide range of deviations from the stoichiometry. However, a single-phase $L2_1$ material is reported only for compositions ranging from Rh$_2$Mn$_{1.32}$Sb$_{0.68}$ to Rh$_2$Mn$_{1.1}$Sb$_{0.9}$ [17,18] but not for stoichiometry. Moreover, several Rh$_2$-based compounds have been reported to crystallize in a tetragonally deformed form of the Heusler structure [19]. A number of tetragonal Rh-based Heusler compounds have magnetic and electrical characteristics that make them very promising for applications involving spin-transfer torque [20]. Several theoretical works have focused on Rh-Mn-Sb due to the generally high interest in Heusler materials [21–26].

It will have been several decades since the last experimental research was pursued with bulk Rh-Mn-Sb alloys, and thin film materials have never been investigated. Here, we present an experimental study of the full Heusler alloy of Rh$_2$MnSb thin film epitaxially grown on MgO (001) in a high purity environment and conditions. A variety of analytical techniques were used to investigate structure, morphology, and phase composition. Optical measurement, Kerr effect, temperature-resolved SQUID, and VSM were used to assess the magnetic and magneto-optical properties.

## 2. Experimental

For every deposition experiment, the following settings remained constant: base pressure below $10^{-7}$ Pa, a target-sample distance of 10 cm, and sputtering Ar gas flow of 14 sccm at a pressure of 2 Pa. Using a Rh$_2$MnSb target, we managed to prepare almost/nearly perfect stoichiometric samples. The target was not a real compound; nonetheless, the target's overall molar ratio of 50% Rh, 25% Mn, and 25% Sb was verified.

By using DC magnetron sputtering, the films were prepared at temperatures ranging from ambient temperature to 800 °C. As the main substrate material, 1 cm x 1 cm × 0.05 cm of surface-polished MgO [100] from MaTeck was utilized. A ratio pyrometer and a thermocouple fastened to the sample stage were used to measure the temperature. An average deposition experiment lasting 30 minutes achieved a layer thickness of 200 +/- 20 nm. The sample parameters are listed in Table 1. The composition and homogeneity of the film were determined by SEM-EDX (Tescan FERA 3). The compositions are in Table 1.

Surface morphology and R$_{MS}$ roughness was revealed by Atomic Force Microscopy (AFM) performed in room temperature mode using ScanAsyst Air tips (Bruker; k=0.4N/m; nominal tip radius 2 nm) in Peak Force Tapping mode on an ambient AFM (Bruker, Dimension Icon). The resolution of the measured topography is $512 \times 512$ points$^2$.

X-Ray measurements (XRD) were performed on X'Pert Pro PANalytical diffractometer with a Co Kα X-ray radiation using Bragg–Brentano geometry with point focus (0.25 mm/0.5 mm divergent slit/mask) and an X'Celerator position-sensitive detector. The sample was attached to ATC-3 texture cradle to measure different patterns in a skew-symmetric geometry. Phase identification was determined by X'Pert HighScore Plus software with PDF-5 database.

Transmission electron microscopy (TEM) was carried out on a FEI Tecnai TF20 X-twin microscope operated at 200 kV. TEM is equipped with an EDAX Energy Dispersive X-ray (EDX) detector, and images were recorded on a Gatan CCD camera with a resolution of 2048 × 2048 pixels using the Digital Micrograph software package. The lamella for TEM observation was prepared using a focused-ion-beam lift-out technique within SEM in a FEI Quanta 3D.

The state of the film was characterized by X-ray Photoelectron Spectroscopy (XPS). The details of used instrument NanoEsca can be found in [27]. The deposited samples were transferred to XPS chamber employing a vacuum suitcase transport [28,29]. The vacuum suitcase transport enabled us to minimize undesirable surface degradation occurring during the time period between the moment the sample is made and the moment the sample is placed in the XPS chamber.

The optical properties of investigated samples were measured at room temperature by Mueller matrix spectroscopic ellipsometer Woollam RC 2. The incident angles of light ranged from 60 to 75 degrees to acquire enough data for subsequent fitting. This was done using proprietary software Woollam CompleteEASE.

Magneto-optical properties were measured at room temperature using a homemade magneto-optical spectrometer based on a rotating analyzer technique in the spectral range from 1.3 to 5 eV. The applied magnetic field was oriented perpendicularly to the sample surface (polar configuration), and the angle of light incidence was nearly normal. The applied magnetic field ranged from 0 to 1 T.

A vibrating sample magnetometer (VSM) of the PPMS (Physical Property Measurement System, Quantum Design Inc.) system was used to measure the magnetic moment of the samples as a function of temperature and magnetic field. The Curie temperature was measured using the temperature dependence of magnetization at a low magnetic field m0H = 0.01 T, which was measured at a rate of 4 K/min. There were two rates of measurement for magnetic hysteresis loops: 1 and 20 mT/s. The minimal available field sweep 1 mT in low field was used to determine the coercive field. Using the estimated density of 8000 kg/m$^3$ and the measured sample dimensions, magnetization was computed from the measured magnetic moment.

## 3. Results and discussion
### 3.1. Sample preparation

Several batches of samples were prepared by deposition at different temperatures and times. The main goal of this stage was to find the optimal temperature and thickness for the preparation of films. It turned out that films with a thickness of around 200 nm show a tile-like structure. Next, the experiment was narrowed down to changing the deposition temperature only. As a result, a batch of samples was finally deposited at temperatures of 600-800°C on MgO exhibiting epitaxial growth. In this work, only these were analyzed. Table 1 summarizes the samples obtained under the conditions described above. We also prepared clone samples to ensure that repeatable results were obtained (denoted further as S600-c).

The average chemical composition of the samples prepared at 600 °C is listed in Table 1, which was obtained by measurement of multiple areas in each sample by EDS analysis. The determined composition $Rh_{2.04}Mn_{1.12}Sb_{0.88}$ is on the edge of the bulk composition with the $L2_1$ ordered structure, and the observed deviation is consistent with the literature [17,18]. The composition's determination error can be about 1 *at.*% [30].

**Table 1.** Summary of sample types prepared in this study. The lattice constants were determined by XRD and $R_{MS}$ by AFM.

| Film notation | Time of deposition [min] | Nominal thickness [nm] | Deposition temperature [°C] | $a$ [nm] | $c$ [nm] | $R_{MS}$ [nm] |
|---|---|---|---|---|---|---|
| **S600** | 30 | 200 | 600 | **0.5884** | **0.6983** | 0.6 |
| **S700** | 30 | 200 | 700 | 0.5887 | 0.6969 | 2.7 |
| **S800** | 30 | 200 | 800 | 0.5916 | 0.6898 | 6.6 |
| Lattice parameters of the bulk alloy: $a$ = **0.5914 nm**; $c$ = **0.6934 nm** [16] (calculated from the cubic-derived coordinates for $Rh_{1.0}Mn_{0.55}Sb_{0.45}$) $a$ = **0.5898 nm**; $c$ = **0.6987 nm** [17] (for the bulk alloy of $Rh_2MnSb$) | | | | | | |
| **The average composition of the sample was prepared at 600 °C, nominally T.** | | | | | | |
| **Sample S600-c** | **Rh (*at.%*)** | | **Mn (*at.%*)** | | **Sb (*at.%*)** | |
| SEM-EDS | 51 | | 28 | | 22 | |
| STEM-EDS | 51 | | 27 | | 23 | |

### 3.2. Microstructure

Fig. 1. shows the morphology of the surfaces of thin films deposited at different substrate temperatures. All films exhibit tile-like morphology with tile edges along [110] direction of MgO substrate. The surface morphology observed in Fig. 1(a) can be attributed to ferroelastic orderings, specifically twinning [31]. A pronounced twinning (Fig. 1 (a)) is well visible by thin film deposited at 600 °C. In this case the difference in height on the surface of this sample does not exceed 4 nm, which indicates a well-structured film with extremely low roughness. Thanks to low roughness, the tiling effect is well-pronounced. Higher deposition temperatures (Fig. 1 (b) and (c)) increase the degree of non-uniformity of the film surface, significantly increasing the height difference and overall roughness (Table 1). This suppresses the visibility of tiling. Although less visible, it is apparent that the character of tiling is very similar to that found in film prepared at 600 °C.

Spontaneous martensitic transformations during thin film deposition produce multiple, regularly oriented twin variants [32]. Like Ni-Mn-Ga, the $Rh_2MnSb$ epitaxial thin films seem to form interrelated 3D hierarchical patterns of twin boundaries in the martensitic phase. The twinning is formed to relieve shear stress from symmetry reduction during the phase transition from the cubic austenitic to the tetragonal martensitic phase. The twin boundary symmetry operators are rotations and/or mirrors [33,34]. Usually, this hierarchical self-accommodation spans a range from the nanoscale to the macroscale [35]. In our case, thin films undergo a martensitic phase transition induced by cooling from high deposition temperatures. In addition, the stress induced during cooling due to substrate constraints can facilitate the transition to a tetragonal state [36]. For $Rh_2MnSb$ thin films, there seem to be only two ferroelastic domains or twin variants in which the in-plane crystal axes are perpendicular to each other, the so-called X-type twin boundaries following the best match on the substrate. The X-type microstructure has a pronounced mosaic surface that height-contrast AFM measurements can see [37].

A weak twinning contrast is also observed in the films prepared at higher deposition temperatures (700 °C and 800 °C). The higher temperature may result in clustering, forming individual clusters of grains on the surface and increasing the roughness. This, in turn, decreases the contrast in twinning morphology or tiling.

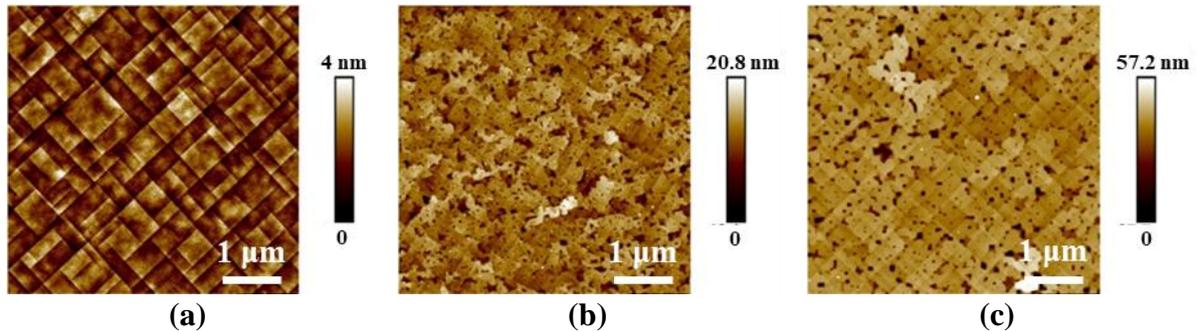

**(a)**            **(b)**            **(c)**

**Fig. 1.** AFM figures of the films prepared in different temperatures: (a) 600 °C; (b) 700 °C; (c) 800 °C

The expected twinning geometry of tetragonal martensite in the film attached to the substrate is schematically sketched in Fig. 2. The figure shows that the twin variants with c-axis perpendicular to surface are a major part of the film and the twin variants with a-axis perpendicular are very narrow forming the thin lines visible on AFM, Fig. 1. This unequal distribution of the twins occurred due to substrate constrain nearly matching diagonal of MgO unit cell to tetragonal lattice constant $a$ of the film. The suggested arrangement in Fig. 2 was confirmed and fully analyzed by TEM and XRD, as described below.

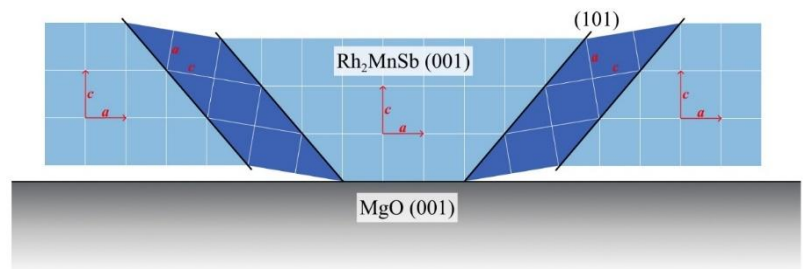

**Fig. 2.** Schematic representation of the twinned system in $Rh_2MnSb$

The comparison between AFM and SEM for another film prepared at 600 °C is shown on Fig. 3. The surface topography of the film (Fig. 3(b)) by SEM validates the findings from the AFM analysis. No further details are revealed by SEM. Moreover, the comparison of the films from Fig. 1(a) and Fig. 3(a) indicates a highly repeatable pattern of the films prepared at 600 °C.

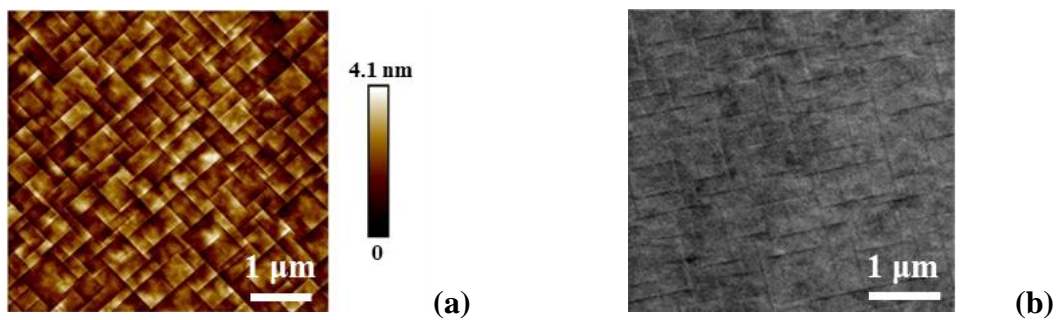

**(a)**            **(b)**

**Fig. 3.** Surface morphology of sample **S600-c**. Comparison of AFM (a) and SEM (b) images

To study the twinning in detail, we prepared cross-sectioned TEM lamellas. Fig. 4 shows results for Sample S600-c, confirming the presence of X-type configurations. Cross-sectional images (Fig. 4(b)-(d)) display alternating bright and dark stripes, tilted at approximately 47°- 50° concerning the substrate – characteristic of an X-type configuration. Here, the twin boundary orientation causes the projections of mirror and rotation symmetry operators in the lamella plane to be equal. The higher resolution TEM in Fig. 4d indicates that the width of the a-variant is very narrow compared to the major twins as sketched in Fig. 2.

Fig. 4 also indicates spatial variations in twinning periodicity ($\Lambda$, ~200 nm). In particular, the periodicity $\Lambda$ does not decrease either near the substrate for twin boundaries or the film surface. Low-period twinning boundaries were also observed at ridge positions that do not reach the surface. In contrast, boundaries at valley positions do not get the substrate because they meet and stop growing before extending throughout the film thickness. More symmetric conjugation interfaces appear at surfaces than at valleys. In $Rh_2MnSb$ films, this could happen if twins nucleate at the same point or grow symmetrically toward each other and are limited by surface constraint.

The TEM observation also concurs with the morphology observed in AFM images; a remarkably smooth, compact, and continuous film is formed without foreign particles or formations. X-ray and electron diffraction structural analyses support the morphology results showing epitaxially grown material. STEM-EDS quantification measurements support previous SEM-EDS results (Table 1). The obtained results demonstrate minor violations of the stoichiometry of the compound.

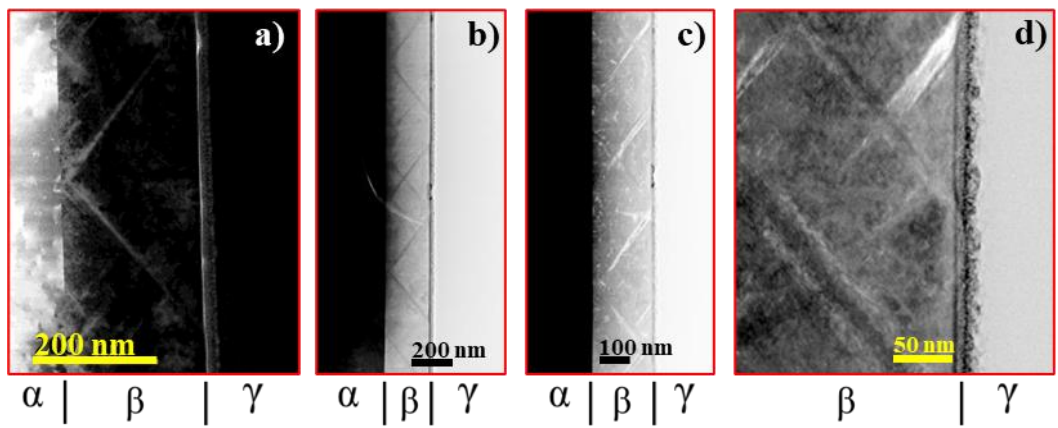

**Fig. 4.** TEM cross-section images of stoichiometric film prepared at 600 °C: α marks the substrate region, β marks the film region, and γ marks the tungsten cover region. a) BF TEM image - the film is oriented into its Zone Axis (ZA) [1-12], b) STEM HAADF image – mixed Z and diffraction contrast – camera length 220 mm, c)-d) STEM HAADF images – camera length 550 mm – predominant diffraction contrast

### 3.3. Crystal structure and twinning

X-ray measurements determined the structure and confirmed the film's expected twinned microstructure. $Rh_2MnSb$ thin films grow epitaxially on MgO with their lattice rotated 45° azimuthally relative to the substrate, stabilizing the following orientation relationships: $(001)[110]_{Rh_2MnSb} \parallel (001)[100]_{MgO}$ and Fig. 5 (a) shows the symmetric ω-2θ scans of the films prepared at different temperatures. It can be noticed that the peaks are shifted due to their different lattice parameters. Furthermore, the lack of additional peaks strongly indicates that Rh$_2$MnSb is the only phase present. Moreover, the presence of superstructure (002) reflection indicates at least $B2$ ordering. To confirm $L2_1$ order other reflections must be measured.

We used skew-symmetric scans to obtain other reflections such superstructure (111) and primary (022) of the epitaxial film. The example for the sample grown at 600 °C is shown in Fig.5(b), which represents all temperatures, as the only insignificant differences observed were minor peak shifts caused by variations in lattice parameters. The peak positions from these measurements were used in a self-built macro, which determines the lattice parameters through a least-squares fit. The calculated lattice parameters are shown in Table 1. The values for the sample S600 are very close to the values from bulk material with similar stoichiometry [16,17]. Minor deviations can be associated with the stress arising from substrate constraints. The volume of unit cells in all films are the same within measuring error. The existence of superstructure reflections (111), (002), (222), and (333) and their magnitude indicates the $L2_1$ structural ordering. Similar diffractograms were obtained for the films prepared in higher temperatures.

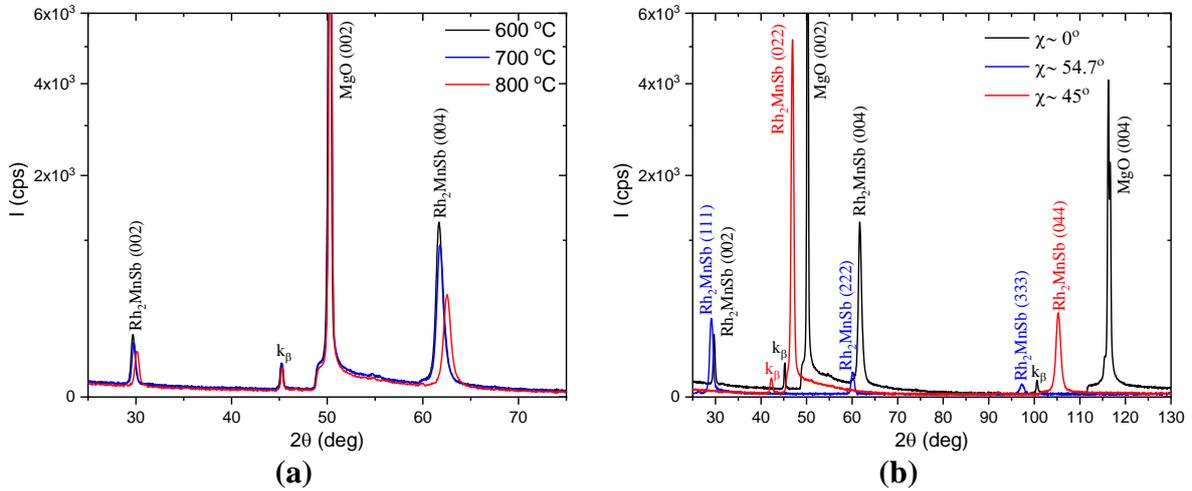

**Fig. 5.** a) X-ray diffraction patterns of the films deposited at 600–800 °C; b) symmetric (002) (in black) and skew-symmetric (111) and (022) (in blue and red, respectively) scans for the sample grown at 600 °C. Vertical axes are truncated for better visualization

Pole figures were measured to study the twinning formation. Since the results are consistent across all three samples made at 600°C, 700°C, and 800°C, we present only the data for the 600 °C sample as a representative example. Fig. 6 (a) shows the 004-pole figure. There is a unique central pole that corresponds to the symmetric (004) reflection, i.e., the c axis is invariantly perpendicularly to the substrate and lattice unit cell parallel to the substrate, as sketched in Fig. 2. On the twinned domains, the *a*-axis is not parallel any more to the substrate, confirmed by the measurement of the 400-pole figure shown in Fig. 6 (b). The presence of four spots originates from the four twin domains, which are symmetrically distributed azimuthally. The expected angle from the geometry of tetragonal twinning is 90 - 2×*arctg*(*c*/*a*) = 9.76° which well correspond to measured angle ~9.3°. Moreover, the intensities of peaks in pole figures for (004) and (400) planes are very different, the difference is more than a thousand times. This reflects the presence of major variants with *c*-axis perpendicular to surface (and *a*-axes in plane) and very small fraction of minor variants lining thinly the major variant domains (Fig. 1). The estimated thickness of minor variant can be close to few unit cells as drawn in

Fig. 2. Similar differences in the diffraction intensities were observed for (022) and (220) pole figures.

Fig. 6 (c) displays the 022-pole figure, revealing four additional poles at χ ≈ 31°, in addition to the expected poles at χ ≈ 50° for a tetragonal structure. These additional spots arise from four twin domains, each rotated by 90° azimuthally relative to each other. In the fourth quadrant, the pole from the main structure is highlighted with a blue circle, while the corresponding pole from one twin domain is marked in red. For further clarity, we provide a sketch of the twinned system in Fig. 7(a), using the same color scheme (main structure in blue and twinned domains in red), along with θ-2θ scans performed on the marked poles mentioned before. Fig. 6(d) presents the pole figure for the (220) reflection. As these planes are perpendicular to the surface, no poles would typically appear here. However, two additional reflections appear in each quadrant due to the presence of twins. The center of this pole figure has been suppressed to avoid the intense (002) reflection tail from the substrate.

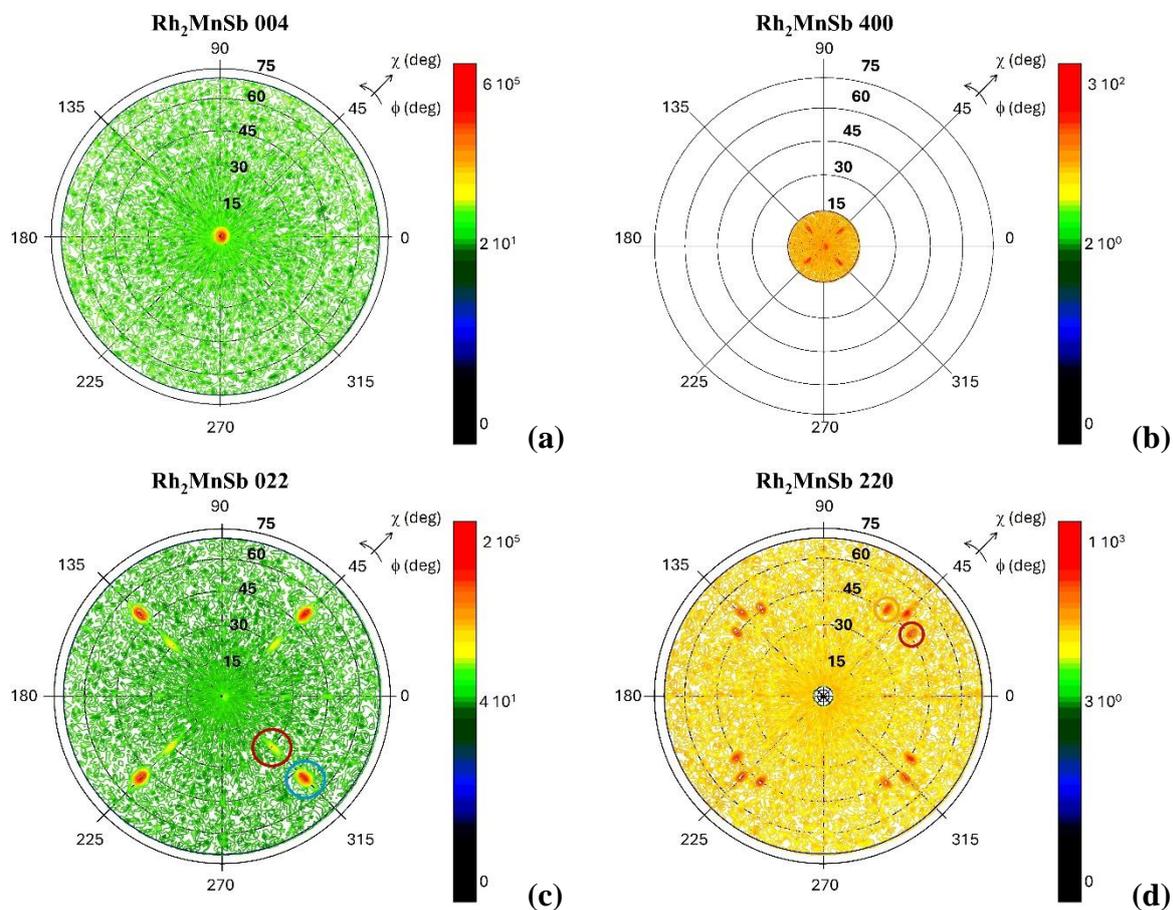

**Fig. 6. (a-d)** Pole figures of films prepared at 600 °C for the main reflection marked in the figure, the φ range is from 0° to 360°, χ range is from 0° to 70.5°

The origin of the eight additional peaks (two of which are highlighted with red and orange circles) is illustrated in the sketch in Fig. 7 (b), along with the diffraction patterns measured at the highlighted spots. Please note that poles lying exactly at φ = 45°, 135°, 225°, and 315° correspond to the right tails of the substrate's (111) peaks and do not belong to the film.

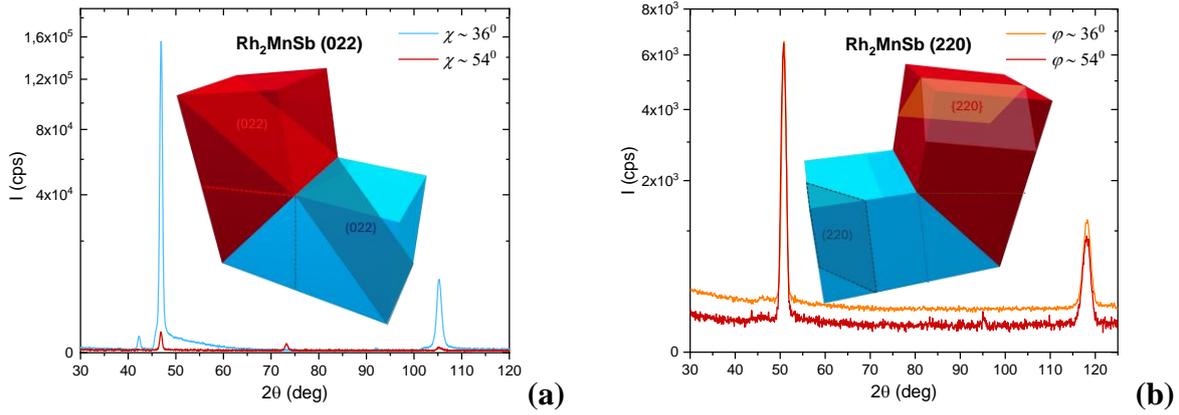

**Fig. 7. (a-b)** Representation of the twinned system and the corresponding θ-2θ scans performed at the φ- χ positions marked by circles in Fig. 6 (c-d).

### 3.4. X-Ray Photoelectron Spectroscopy

To check for the presence of oxides and to investigate electronic state of material XPS analysis was performed on a vacuum suitcase transported sample S600. Of the metals constituting the Rh$_2$MnSb Heusler, Mn is the most sensitive to the ambient environment [38]. On the other hand, Rh is a noble metal which does not readily form native oxides [39]. Measured spectra's are presented in Fig. 8. During measurements, the charging of the sample occurred. To cope with the charging, it was important first to inspect the Sb 3d and O 1s spectral region for presence of oxygen. The spectral overlap of Sb 3d and O 1s represents a complication but it is possible to deduce on the presence of oxygen by comparing peak areas of the Sb 3d doublet [40]. A ratio larger than $3/2$ would imply oxygen presence. However, the measured area ratio corresponds to the correct $3/2$ thus excluding oxygen presence. As there was no obvious oxygen contribution in the spectra, we assumed that Rh present in our sample is indeed in metallic form and so we used its Rh 3d $5/2$ peak for charging correction.

As stated above it was not possible to find an oxygen contribution. Also, there was no carbon contribution. This implies high sample purity. Comparing with the literature reported peak positions, there is a very good agreement with metallic state for all the constituting metals [40,41]. Further, Mn 2p was particular with a typical high peak asymmetry. The asymmetry was lower for Sb and Rh. However, a special spectral feature was observed in Mn 2p and Sb 3d spectra. There were peak components at higher binding energies that, at first look, would suggest oxide bonds. In fact, for Mn it can be ascribed to originate from a 2p-3d exchange interaction resulting in splitting of the peak into two subpeaks separated by around 0.9 eV. This phenomenon has already been observed and described for the Mn 2p core level [27,42–44], typical for Heusler alloys with Mn. Here we observed the well-defined magnetic splitting of the Mn 2p $3/2$, 2p $1/2$ lines in XPS spectra. This gives direct evidence of the existence of well-defined localized magnetic moments on Mn in comparison with itinerant-electron ferromagnets. Such localization is typical for Heusler alloys. Surprisingly the splitting of Sb 3d was also observed although no appreciable magnetic moment is expected on Sb atoms in a Heusler structure. No splitting was found for Rh. Further investigation of magnetic moment distribution is clearly needed.

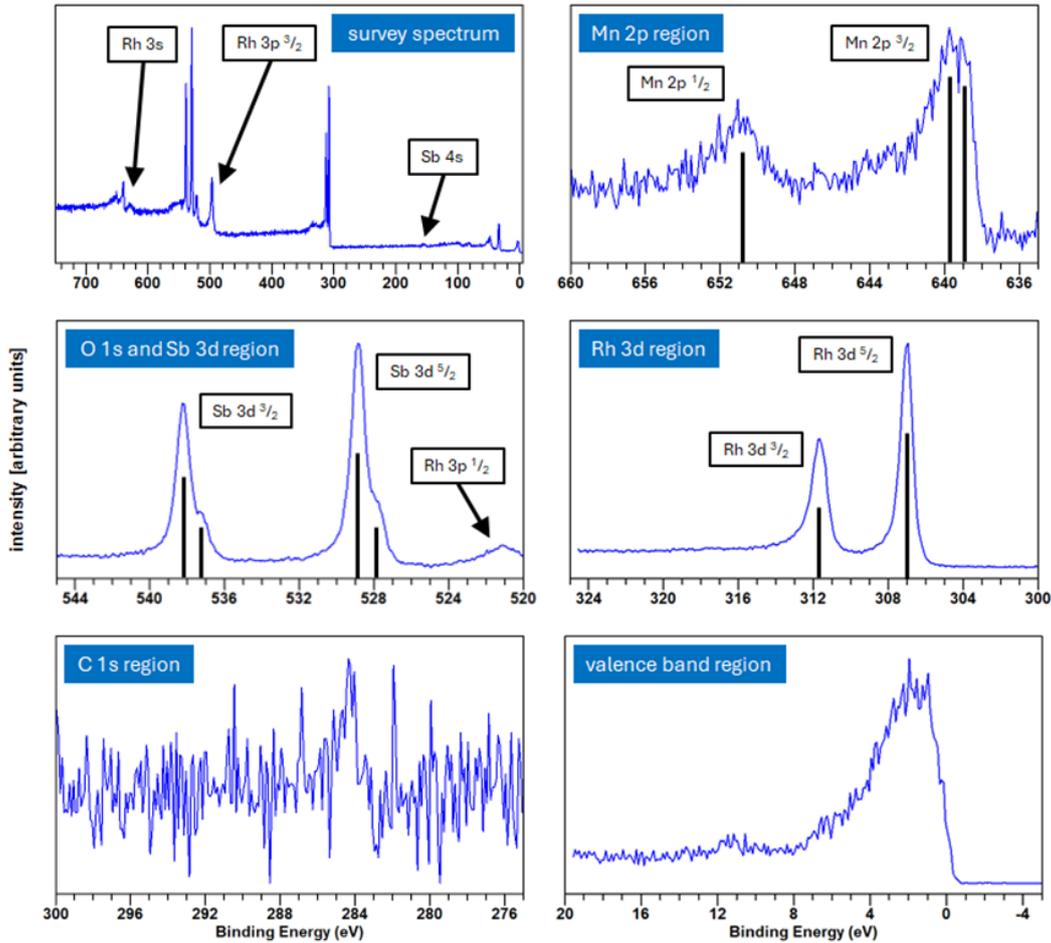

**Fig. 8.** XPS spectra of vacuum transported freshly prepared S600 type sample. A low-resolution survey spectrum as well as high resolution spectra centered around spectral regions of respective elements are presented. Peak labeling for binding energies lower than 100 eV is omitted for clarity. Literature reported value of Rh 3d $^{5}/_{2}$ in metallic state [40,41] was used for charging correction. Black vertical lines denote significant peak positions: 651.0, 639.6, 638.7 eV = metallic Mn; 538.2, 537.3, 528.8, 527.9 eV = metallic Sb; 311.7, 307.0 eV = metallic Rh. Mn 2p and Sb 3d display a particular ~ 0.9 eV splitting of the metallic peak.

### 3.5. Optical and magneto-optical properties

Concerning the surface quality shown by AFM measurements, we limit ourselves to the investigation of optical and magneto-optical properties of the sample S600, which exhibited the lowest surface roughness and visible twining.

Obtained experimental ellipsometric parameters PSI and DELTA were fitted using a model structure of semi-infinite MgO substrate with a 200 nm thick $Rh_2MnSb$ layer on top. A surface roughness of 0.6 nm was taken from AFM measurements and included in the model as well. A BSpline function was used to parametrize the spectral dependence of the complex permittivity of the $Rh_2MnSb$ layer. The resulting spectral dependence of real and imaginary parts of permittivity is shown on Fig. 9. The spectral behavior resembles that of other Heusler compounds, such as $Co_2TiSn$ [29], $Co_2MnSi$ [45], or $Co_2FeAl_{0.5}Si_{0.5}$ [46] with a rapid increase of Im(ε) towards lower energies, which is typical for metallic systems and intra-band optical transitions. On the other hand, the spectrum of Re(ε) exhibits several spectral structures located near 1.1, 2.1, 3.5, and 4.5 eV, suggesting the involvement of several electronic transitions originating from Mn and Rh *d* electrons [47].

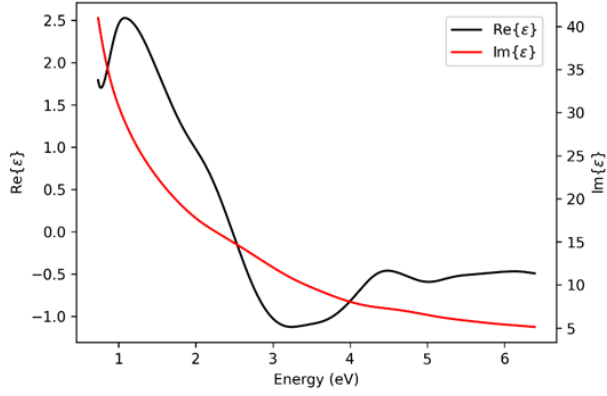

**Fig. 9.** Spectral dependence of the real and imaginary parts of permittivity for the sample S600

Experimental spectra of magneto-optical polar Kerr rotation as a function of the applied external magnetic field are shown in Fig. 10 (a). The amplitude of the polar Kerr rotation scaled linearly with the applied magnetic field (Fig. 10 (b) for the energy of 3.5 eV), which confirmed the paramagnetic behavior of the Rh$_2$MnSb layer at room temperature. A slight offset between up and down branches in Fig. 10 (b) was lower than 0.5 mdeg, which was below the experimental error of the measurement system. This offset might be induced by the magnetic hysteresis of the magnetic circuit used in the electromagnet to generate the external magnetic field. The amplitude of the polar Kerr rotation spectrum for an external magnetic field of 1T reached its maximal value of approx. 12 mdeg at lower energies. Such value was far below the polar Kerr rotation amplitudes of ferromagnetic Heusler compounds containing manganese [45], again reflecting the $T_C$ below room temperature and no magnetic order. The spectral behavior, however, is typical for manganese compounds with a notable negative spectral structure located around 3.5 eV, followed by a positive spectroscopic structure near 1.2 eV. These values agree with the Re(ε) spectral behavior shown in Fig 9. Similar trends were observed, for example, in Ni-Mn-Ga [48,49] in the martensitic phase. One can, therefore, consider Rh$_2$MnSb electronic structure with half-metallic behavior as in Ni-Mn-Ga. Theoretical calculations of element-resolved densities of states (DOS) by Benichou et al. showed sharp peaks in DOS of Rh $d$ states at approximately 2 eV bellow Fermi energy and of Mn $d$ states at approximately 1.5 eV above Fermi energy [47], suggesting that the main spectral features in polar Kerr rotation more likely originate from Rh to Mn $d$ states transitions. However, more detailed theoretical and experimental studies at low temperature would be necessary to ascribe all the transitions to experimental spectra.

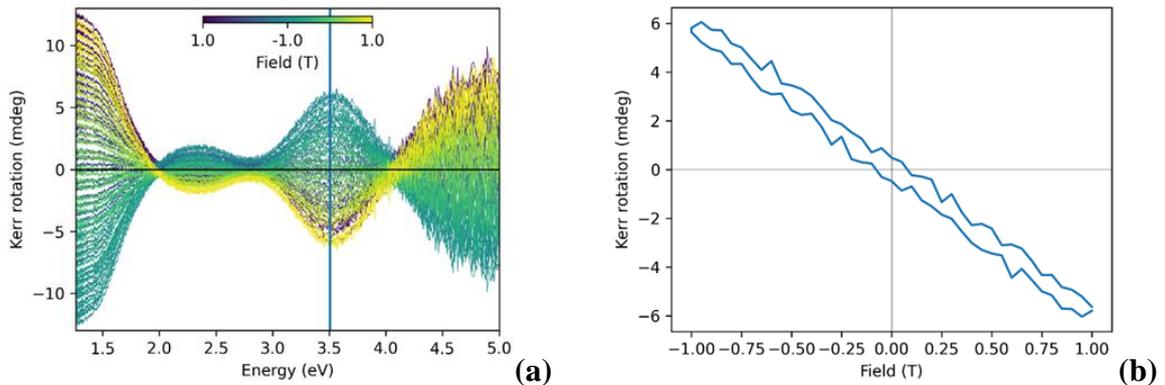

**Fig. 10. a)** Exp. spectra of magneto-optical polar Kerr rotation as a function of external magnetic field. **b)** magneto-optical polar Kerr rotation taken at 3.5 eV as a function of applied magnetic field

## 3.6. Magnetic properties

Magnetic properties were measured up to 9T in the range 10-400 $K$. Although several films were measured, the results from only one exemplary film prepared at 600 °C are shown here.

Saturation magnetization $M_S$ at 10 $K$ was about 55 emu/g, and Curie point $T_C$ = 220-275 $K$ depending on the films. The films prepared in higher temperatures exhibit a higher Curie point of about 260 $K$, which may be ascribed to minor composition change. For comparison, the value of $M_S$ is close to the value of a bulk alloy with a similar stoichiometry (52.3 emu/g at absolute zero [16]), while the $T_C$ of bulk alloys is higher (304 $K$ [17]; 330 $K$ [16]; 348 $K$ [19], probably again reflecting slight changes of compositions). Another possible reason for the lower Curie points in the obtained thin films is reduced dimensionality and surface effects, which weaken magnetic interactions. In addition, stresses from substrate constrain and twinning may also affect the transition temperature. The saturation magnetization is low compared to other stoichiometric Heusler alloys, indicating that the excess Mn is arranged antiferromagnetically, similar to Ni-Mn-Ga [50] and, in contrast to $Ni_2MnGa$ where Ni is magnetic, the Rh had no magnetic moment as indicated by XPS.

Fig. 11 (a) shows that the thermomagnetic curve is smooth without visible jumps, indicating no structural transformation at this temperature interval. The Curie temperature was determined from a sharp decrease of low field magnetization. Moreover, there was a weak tail up to 350 $K$. This ferromagnetic tail might suggest the presence of another magnetic phase, but no such phase was indicated by XRD or electron microscopy. The observed magnetization tail can originate from the twin boundaries, which is expected to have different structures and thus different magnetic arrangements compared to materials inside the twin domain, and therefore, it may exhibit different Curie temperature.

The magnetization curves were measured in-plane and along the edge of the MgO substrate, e.g., along [110] direction of tetragonal martensite. In high field the curves were strongly affected by diamagnetic contribution from MgO substrate. After the diamagnetic substrate contribution was subtracted the curves up to 0.2 T are shown in Fig. 11(b) to resolve details of magnetization curves and hysteresis. At a temperature just below Curie point, hysteresis was very low ($\approx 15 \times 10^{-3}$ T) and increased with decreasing temperature up to ($\approx 61 \times 10^{-3}$ T) at 10 $K$, indicating increased magnetic anisotropy. This increase is also apparent by the gradual tilting of the magnetization loop. However, the in-plane anisotropy is weak compared to the tetragonal martensite of, e.g., Ni-Mn-Ga [43]. The value estimated from the tilt or anisotropy field is about $5 \times 10^4$ J/m$^3$ at 10 K. One reason for low anisotropy can be non-magnetic character of Rh atoms in contrast to magnetic character of Ni in Ni-Mn-Ga which determines the anisotropy [51]. A very small but apparent magnetization loop was measured at 300 K correlating with the magnetization tail in thermomagnetic measurement.

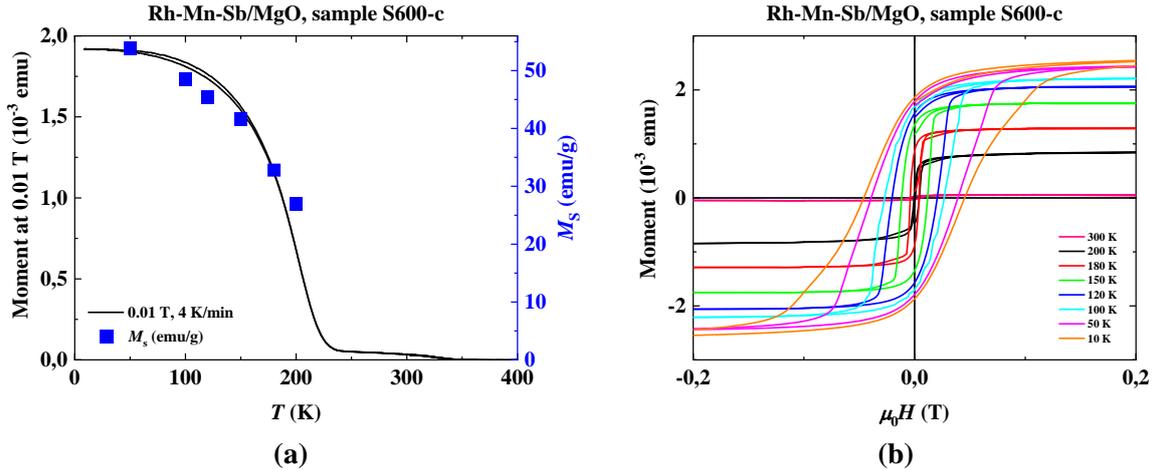

**Fig. 11.** **(a)** Thermomagnetic curves measured at 0.01 T, and **(b)** temperature dependence of saturation magnetization obtained from magnetization curves measured in-plane up to 9T after correction for the diamagnetic substrate.

## 4. Conclusion

We have deposited thin films of the half-metallic material $Rh_2MnSb$ by DC magnetron sputtering, which grows in a single phase. The structural, compositional, magneto-optical, and magnetic properties of $Rh_2MnSb$ thin films were studied. The grown films are slightly off-stoichiometric, with a minor excess of Rh and Mn. The depositions were performed under high purity conditions which positively reflected in the absence of non-metallic contamination. The desired $L2_1$ ordering particular to Heusler alloys was confirmed. The lattice parameters for the tetragonal $Rh_2MnSb$ unit cell were close to the bulk alloy. The film obtained at 600 °C exhibit very smooth surface and all films revealed a regular twinning, which was studied in detail. The polar Kerr rotation spectra for a film at 600 °C reflects paramagnetic behavior and $T_C$ below room temperature without magnetic order. Magnetic measurements found that this film has $T_C \approx 220\text{-}275\ K$ and saturation magnetization of 55 emu/g at 10 K, similar to bulk alloy. Due to the half-metallic behavior, twins and regular twin interfaces, $Rh_2MnSb$ thin films may be interesting candidates for spintronic applications and can be used as a shape memory material.

## Acknowledgment

We acknowledge the support provided by the Ferroic Multifunctionalities project, supported by the Ministry of Education, Youth, and Sports of the Czech Republic [Project No. CZ.02.01.01/00/22_008/0004591] co-funded by European Union and Czech Science Foundation grant No. 24–11361S. Magnetic measurements were performed in MGML (http://mgml.eu), which is supported within the program of Czech Research Infrastructures (project No. LM2023065). The authors would like to thank Ladislav Klimša, Jarmila Balogová, and Petr Svora from the Institute of Physics of the Czech Academy of Sciences for their assistance in sample preparation and SEM measurements.